\documentclass[journal]{IEEEtran}

\ifCLASSINFOpdf
\else
   \usepackage[dvips]{graphicx}
\fi
\usepackage{url}

\usepackage{amsmath}
\usepackage{amssymb}
\usepackage{amsfonts}
\usepackage{graphicx}
\usepackage[dvipsnames]{xcolor}
\usepackage{cite}
\usepackage{rotating}
\usepackage{url}
\usepackage[ruled,vlined]{algorithm2e}

\begin{document}

\title{Sparse Antenna Array Design for MIMO Radar Using Softmax Selection}

\author{Konstantinos Diamantaras, \IEEEmembership{Senior Member, IEEE}, %
Zhaoyi Xu, \IEEEmembership{Student Member, IEEE}, and %
Athina Petropulu, \IEEEmembership{Fellow, IEEE}%
\thanks{Work supported by NSF under grant ECSS-2033433.}%
\thanks{K. Diamantaras is with the Department of Information and Electronic Engineering , International Hellenic University, Sindos, 57400, Thessaloniki, Greece, E-mail: \texttt{k.diamantaras@ihu.edu.gr}.}%
\thanks{Z. Xu and A. Petropulu, are with the Department of Electrical and Computer Engineering, Rutgers University, Piscataway, NJ, USA, E-mail: \texttt{\{zhaoyi.xu,athinap\}@rutgers.edu}}}

\markboth{Submitted to IEEE Signal Processing Letters}{Diamantaras \MakeLowercase{\textit{et al.}}: Sparse Antenna Array Design for MIMO Radar}

\maketitle

\begin{abstract}
MIMO transmit arrays allow for flexible  design of the transmit beampattern. However, the large number of elements required to achieve certain performance using  uniform linear arrays (ULA) maybe be too costly. This motivated the need for thinned arrays by appropriately selecting a small number of elements so that the full array beampattern is preserved. In this paper we propose \textit{Learn-to-Select} (\textit{L2S}), a novel machine learning model for selecting antennas from a dense ULA employing a combination of multiple Softmax layers constrained by an orthogonalization criterion. The proposed approach can be efficiently scaled for larger problems as it avoids the combinatorial explosion of the selection problem. It also offers a flexible array design framework as the selection problem can be easily formulated for any metric.
\end{abstract}

\begin{IEEEkeywords}
MIMO radar, sparse array, antenna selection,  machine learning, softmax
\end{IEEEkeywords}

\IEEEpeerreviewmaketitle

\section{Introduction}

\IEEEPARstart{M}IMO transmit arrays allow for flexible  design of the transmit beampattern. 
In the  presence of interference, an optimally designed beam  can improve target detection in radar, and increase throughput or channel capacity in MIMO wireless communication systems. However, while antennas are inexpensive, the antenna RF fronts are expensive, and using as few RF fronts as possible is  desirable. Sparse arrays, also referred to as thinned arrays offer a reduced cost approach,  and can be formed based on a full uniform array, by activating only a small number of antennas. 
Selecting the active antennas so that the  sparse array has performance close to that of a full array, has been attracting a lot of attention.

Here, we consider a MIMO radar that transmits partially correlated waveforms from its transmit antennas. The waveform diversity allows higher degrees of freedom as compared to phase arrays, which enables the MIMO radar to flexibly design the transmit beampattern.
For
 fixed transmit antenna positions, MIMO radar beampattern design  amounts to finding the   optimal transmit waveform covariance matrix  
 that results in the desirable  beampattern 
 \cite{[12],[14]}.
Receive beampattern design for thinned receive arrays   \cite{[18],[26],[28],[29],[30],[32],[33],[35],[36],[38]}
seeks  to
select the receive antenna positions as well as compute the beampattern weights via various
optimization strategies \cite{[28],[32],[30],[29],[26]}, for example, by  sequential addition of
antennas, such that antenna weights and positions
are optimized at each addition \cite{[26]}. Another approach \cite{[27]} is
to start with a full array structure and then iteratively
remove the antennas which produce the highest sidelobe
levels in the  beampattern. As the problem is in general nonconvex, only locally optimal sparse array design approaches  exist.
 These iterative beampattern design methods were also extended to
 sparse MIMO arrays in \cite{Roberts2011}. 
  Optimal  sparse array design has been approached via global optimization techniques, such as particle swarm optimization \cite{Jin_2007,Schmid_non_uniform_2009,Simoni_optimal_array_ICMIM_2018}.
 In \cite{Hamza-Amin-2019},    sparse array design for receive beamforming based on maximizing the signal-to-interference plus noise ratio is proposed. 
In \cite{Godrich2012}, transmit and receive antenna selection is formulated in a combinatorial optimization framework as a knapsack problem (KP), using the Cramér-Rao bound of the localization error as a performance metric.
In general, the antenna selection schemes that have been proposed in the literature are developed for specific metrics, e.g., minimization of the Cramér-Rao lower bound of the velocity estimates, localization error minimization, interference cancellation and SINR maximization. Those formulations typically result in NP hard optimization problems, which require relaxation techniques and various assumptions in order to reach a solution.

In this paper
we propose
\textit{Learn-to-Select} (\textit{L2S}),
a novel machine learning (ML) approach, to address the exponential complexity of the antenna selection problem. 
Via L2S, we design a sparse linear array by selecting elements from a dense uniform linear array (ULA) using a combination of multiple Softmax neural networks. 
The idea is inspired from the dot-product attention mechanism \cite{Bahdanau2014neural} which has been instrumental in the design of the popular Transformer models \cite{vaswani2017attention}.
Since we need to select $M$ out of $N$ antennas, we employ an equal number of Softmax models. In order to avoid repeating the same solution among the models, we impose an additional orthogonalization constraint.
Our learning model simultaneously optimizes the multiple Softmax selection models and the precoding matrix of the transmitted signals.

Machine Learning approaches for antenna selection have been proposed in \cite{Joung-2016},\cite{Elbir-Mishra-2020},\cite{Vu2019machine}.
In these papers however, the problem is either treated as a classification problem with ${N \choose M}$ classes \cite{Elbir-Mishra-2020}, \cite{Vu2019machine}, or it involves a neural network with a target vector of size ${N \choose M}$ \cite{Joung-2016}. The drawback of these method is the combinatorial explosion problem which renders them impractical in cases with large or even medium numbers of antennas.


%
 %

%
With the use of existing programming tools and parallel platforms (eg. Tensorflow and GPUs) L2S can be easily and efficiently implemented. In addition to that, L2S can be formulated to accommodate any cost function and could be extended to any optimization problem that requires item selection from a list of available items.
%
Because of its flexibility, the proposed L2S has much broader impact that any of the existing methods, as it can be used to solve selection problems arising in many different scenarios.

\section{Problem formulation}

We consider a uniform linear array with $N$ antennas, which are equally spaced with separation distance $d$.
We assume that the antennas transmit narrow-band signals with carrier wavelength $\lambda$.
The array output at angle $\theta$ is
\begin{equation}
    y(t;\theta) = \mathbf{a}(\theta)^H \mathbf{v}(t)
\end{equation}
where $\mathbf{a}(\theta)$ is the steering vector at direction $\theta$, and $\mathbf{v}(t) \in \mathbb{C}^N$ is the transmitted array snapshot at time $t$.
Let
\begin{equation}
    \mathbf{v}(t) = \mathbf{Q}\mathbf{e}(t)
\end{equation}
where $\mathbf{e}(t) \in \mathbb{C}^N$ is a white signal vector
and $\mathbf{Q}\in \mathbb{C}^{N\times N}$ is a precoding matrix.
The array output vector at $K$ different angles 
is
\begin{eqnarray}
   \mathbf{y} &\triangleq& [y(t;\theta_1),\dots,y(t;\theta_K)]^T \\
   &=& \mathbf{A}^H \mathbf{v}(t)
\end{eqnarray}
 with
$\mathbf{A}=[\mathbf{a}(\theta_1), \dots, \mathbf{a}(\theta_K)] \in \mathbb{C}^{N\times K}$ being the steering matrix.
We want  to select $M$ array antennas ($M<N$), so that the array achieves the desired power at the $K$ angles.
We approach the problem by defining $\mathbf{S} \in \mathbb{R}^{M\times N}$ to be a sparse selection matrix such that all its elements are zero, except for exactly one element per row which is equal to one. 
In general, if we select positions $l_1,\dots,l_M$ then the elements of $\mathbf{S}$ will be
\begin{equation}
    s_{ij} = \left\{ \begin{array}{rl}
        1 & \text{if } l_i = j\\
        0 & \text{otherwise}
    \end{array}\right.
\end{equation}
Correspondingly,  matrix $\mathbf{S}^T\mathbf{S}$ will be an $N\times N$ diagonal matrix with all diagonal entries equal to zero except those at positions $l_1,\dots,l_M$.

The  output of the sparse array can be expressed as
\begin{eqnarray}
\mathbf{y}_s(t) &\triangleq& [y_s(t;\theta_1), \dots, y_s(t;\theta_K)]^T \\
&=& \mathbf{A}^H\mathbf{S}^T\mathbf{S} \mathbf{v}(t)
\end{eqnarray}



Let $p_i = p(\theta_i)$ be the desirable signal power at angle $\theta_i$, so that the desired beam-pattern is the vector $\mathbf{p}=[p_1,\dots,p_K]^T$.
The sparse array output power at $\theta_i$ is
\begin{eqnarray}
    \hat{p}_i &=& \mathbb{E}\{y_s(t;\theta_i)^* y_s(t;\theta_i)\} \\
    &=&
     \mathbf{a}(\theta_i)^H \mathbf{S}^T \mathbf{S}
    \mathbf{R}_v \mathbf{S}^T \mathbf{S} \mathbf{a}(\theta_i)
\end{eqnarray}
where $\mathbf{R}_v = \mathbb{E}\{\mathbf{v}\mathbf{v}^H\}
 = \mathbf{Q} \mathbf{Q}^H$.
We want to find the selection matrix $\mathbf{S}$ and the precoding matrix $\mathbf{Q}$ that minimize the beam-pattern error
\begin{eqnarray}
    \mathcal{L}(\mathbf{S},\mathbf{Q})
    &=& \sum_{i=1}^K (p_i - \hat{p}_i)^2
    \nonumber \\
    &=& \|\mathbf{p} - \text{diag}\{\mathbf{A}^H  \mathbf{S}^T \mathbf{S}
    \mathbf{Q} \mathbf{Q}^H
    \mathbf{S}^T \mathbf{S} \mathbf{A}\}\|^2
    \label{eq:loss}
\end{eqnarray}

\section{Softmax co-design}

We propose a novel machine learning approach for the co-design of  $\mathbf{S}$ and $\mathbf{Q}$, given the loss function $\mathcal{L}$.
Each of the $M$ rows of $\mathbf{S}$ can be modeled by a separate softmax neural network \cite{bishop2006pattern} with $N$ outputs, i.e., as many as the antenna elements in the original array. The outputs of the $m$-th network will be
\begin{equation}
    s_{m,i} = \frac{\exp(\mathbf{w}_i^T \mathbf{x} + b_i)
    }{\sum_{j=1}^N \exp(\mathbf{w}_j^T \mathbf{x} + b_j)
    }, ~~~~i=1,\dots,N
\end{equation}
where $\mathbf{w}_i$, $b_i$ are respectively the weights and biases, and $\mathbf{x}$ is a the input.
Note that
$
0 \leq s_{m,j}\leq 1
$
and
\begin{equation}
    \sum_{j=1}^{N}s_{m,j} = 1.
    \label{eq:sum_s}
\end{equation}
Essentially, $s_{m,i}$ represents the probability that antenna $i$ will be our $m$-th selected antenna.

Since the selection matrix does not depend on  time $t$, the input $\mathbf{x}$ should be constant, and thus, the constant value $b_i' = \mathbf{w}_i^T \mathbf{x}$ can be absorbed into the bias term $b_i$.
Without loss of generality, such a model is equivalent to a softmax model with $\mathbf{x}=0$ where the only trainable parameters are the biases.

\begin{figure}[htb]
\begin{minipage}[b]{1\linewidth}
  \centering
  \centerline{\includegraphics[width=7cm]{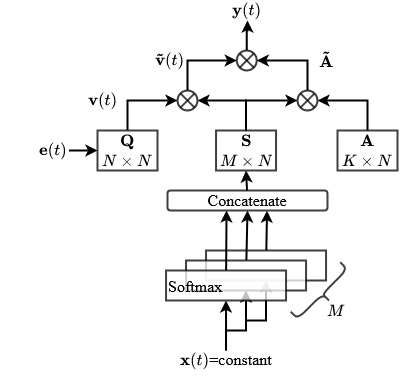}}
\end{minipage}
\caption{Signal flow graph in the \textit{Learn2Select} model. The concatenation of the $M$ Softmax models creates the selection matrix $\mathbf{S}$. The matrix $\mathbf{Q}$ and the Softmax models weights/biases are the parameters to be trained.}
\label{fig:ml_model}
\end{figure}

The approximated selection matrix, $\mathbf{\hat{S}}$
is formed based on the outputs
    $\mathbf{s}_m = [s_{m,1},\dots,s_{m,N}]^T$
of all the softmax models as its columns.
Clearly, $\mathbf{\hat S}$ will be a soft selection matrix since the values $s_{m,i}$ range between 0 and 1.
By the end of the training, the matrix should converge very close to hard binary values so the approximation will be good.
The architecture of the Machine Learning model is shown in Fig. \ref{fig:ml_model}.

{\color{black}
In order to formulate the cost function we express the real and imaginary parts of $y_s(t;\theta_k)$
as functions of the real and imaginary parts of $\mathbf{A}$, $\mathbf{Q}$, $\mathbf{v}$ and $\mathbf{e}$.
This is necessary to facilitate the machine learning optimization which is based on real numbers.
}

The average output power at angle $\theta_k$ is 
\begin{eqnarray}
    \tilde{p}_k &=& \frac{1}{T} \sum_{t=1}^T y_s^*(t;\theta_k) y_s(t;\theta_k)
\end{eqnarray}
so the approximation error with respect to the desired beam-pattern is
\begin{equation}
    \mathcal{\tilde L} = \sum_{k=1}^K \gamma_k (p_k - \tilde{p}_k)^2
\end{equation}
where $\gamma_k$ is the importance weight assigned to the angle $\theta_k$.

In order to achieve a realistic solution, the softmax models must produce hard binary values. The following constraint enforces this requirement:
\begin{equation}
    \sum_{i=1} s_{mi}^2 = 1,
    ~~~~\text{for all }m.
    \label{eq:constraint_1}
\end{equation}
Indeed, \eqref{eq:constraint_1} holds iff $s_{mi} \in \{0, 1\}$.
The `if' part of this statement is obvious.
The `only if' part comes readily from \eqref{eq:sum_s} since
\begin{align*}
\Bigl[\sum_{i=1} s_{mi}\Bigr]^2 - \sum_{i=1} s_{mi}^2 = 0 \\
\Rightarrow 2\sum_{i\neq j} s_{mi} s_{mj} = 0
\end{align*}
implying that at most one element of $\mathbf{s}_m$ can be equal to 1 and all other elements must be equal to 0.
Combined with \eqref{eq:sum_s} this means that \textit{exactly} one element of $\mathbf{s}_m$ is equal to 1 and all other elements are equal to 0.

We also need to impose a further constraint so that the same antenna is not selected more than once, i.e.
\[
    s_{m,i} = 1 \Rightarrow s_{n,i} = 0,
    ~~~~\text{for all } n\neq m
\]
If $s_{m,i} \in \{0,1\}$ then the above constraint is equivalent to
\begin{equation}
    \mathbf{s}_m^T \mathbf{s}_n = 0.
    \label{eq:constraint_2}
\end{equation}
Combining \eqref{eq:constraint_1} and \eqref{eq:constraint_2} it follows that $\mathbf{\hat S}\mathbf{\hat S}^T$ must be equal to the identity matrix $\mathbf{I}$.
Therefore we introduce an extra penalty term in the overall loss function, i.e., 
\begin{equation}
    \mathcal{L}_0(\mathbf{b},\mathbf{Q})
    = \mathcal{\tilde L} + \alpha \|\mathbf{\hat S}\mathbf{\hat S}^T - \mathbf{I}\|_F^2 .
    \label{eq:learning_loss}
\end{equation}
where $\|\cdot\|_F$ denotes the matrix Frobenius norm.
The cost parameter $\alpha$ reflects the relative importance of the latter constraint 
with respect to the desired beam-pattern  error.
The cluster of Softmax models in the L2S model remotely resembles the multi-head attention architecture proposed in \cite{vaswani2017attention}. The major differences are (a) the proposed normalization scheme \eqref{eq:learning_loss} compared to simple scaling (b) concatenation happens directly on the Softmax layers and (c) the query values are zero. Conceptually however, L2S works similarly to the attention mechanism: it tries to identify the important antennas from the sequence of available ones.

\subsection{Learning to select antennas}

There are two sets of parameters to be trained:
(i) the biases $b_1,\dots,b_M$ (assuming $\mathbf{x}=0$), and (ii) the covariance shaping matrix $\mathbf{Q}$.
We propose a two stage optimization approach alternating between fixing $b_i$ and optimizing over $\mathbf{Q}$
   and then fixing $\mathbf{Q}$ and optimizing over $b_i$.

The algorithm runs for $N_{epoch}$ learning epochs and each alternating stage runs for a small number of steps $N_{step}$.
We can use standard gradient descent (GD) optimization for each step with learning rate parameter $\beta$.
The proposed L2S scheme  is shown in Algorithm \ref{alg:learn_to_select}.
Alternatively, other optimization algorithms instead of standard gradient descent can be used in order to improve speed of convergence. In our simulations we have used the Adam optimizer proposed in \cite{Kingma2014adam} which has shown superior performance compared to GD.

\begin{algorithm}[h]
\SetAlgoVlined
\DontPrintSemicolon
\For{epoch=$1$ \KwTo $N_{epochs}$}{
    Fix $\mathbf{Q}$ and optimize $\mathcal{L}_0$ w.r.t. $b_i$:\;
    \For{step=$1$ \KwTo ~$N_{steps}$}{
      Update $b_i, ~~~~ i=1,\dots,M$\;
    }
    Fix $b_i$, $i=1,\dots,M$ and optimize $\mathcal{L}_0$ w.r.t. $\mathbf{Q}$:\;
    \For{step=$1$ \KwTo $N_{steps}$}{
      Update $\mathbf{Q}$\;
    }
}
\caption{Learn to select.}
\label{alg:learn_to_select}
\end{algorithm}

\section{Simulation results}

Our first experiment is designed to select $M=8$ elements from a ULA $N=16$ antennas spaced apart by  
$d=\lambda/2$.
The beam power profile that we want to approximate 
 is equal to $1$ at  angle ranges $[-2,2]$ degrees and $[28,32]$ degrees, and is zero otherwise. 
 The $0$ degree angle is perpendicular to the face of the antenna array.
In all experiments we used a flat weight function $\gamma_k = 1$.
We used the Adam stochastic optimization procedure with fixed learning rate $\beta=0.01$ and $400$ epochs of training. In each epoch $10$ steps are executed.
The algorithm receives a white input sequence $\mathbf{e}(t)$ of length $T=100$. In applying this method
it should hold that $T>N$.
The parameter $\alpha$ used in \eqref{eq:learning_loss} changes between learning epochs;  it starts from  $\alpha_{init} = 5$, and is linearly incremented a to reach the final value $\alpha_{final}= 25$ at the final epoch.

\begin{figure}[htb]
\begin{minipage}[b]{1\linewidth}
  \centering
  \centerline{\includegraphics[width=7.5cm]{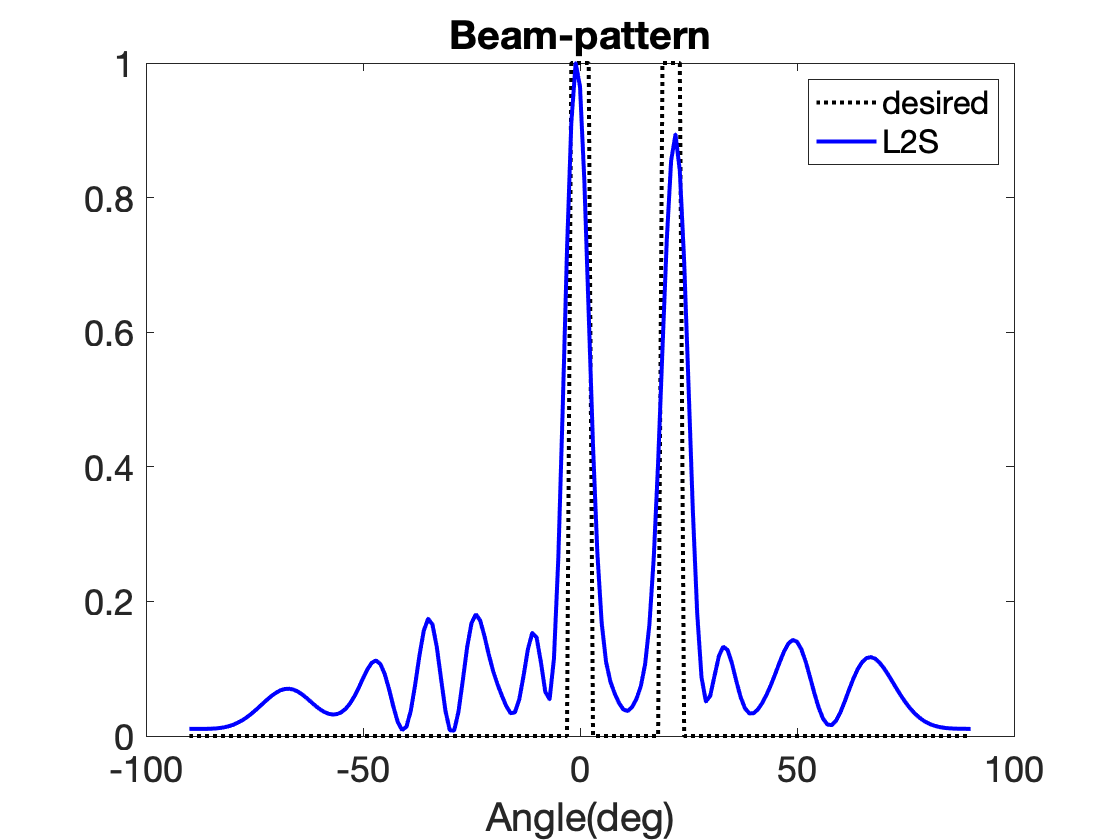}}
  \centerline{\includegraphics[width=9.5cm]{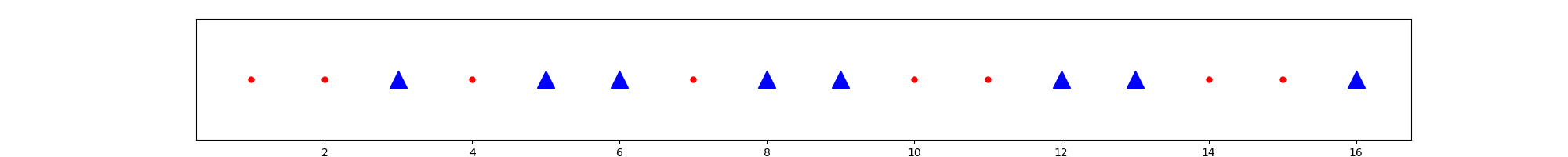}}
\end{minipage}
\caption{(Top) Desired and constructed beam pattern based on $8$ antennas selected from a $16$-element ULA via L2S. (Bottom) Selected antenna locations.}
\label{fig:power2}
\end{figure}

The desired and and approximated beam-patterns are shown in Figure \ref{fig:power2} while the selected antenna locations are shown in Fig. \ref{fig:power2}.
The half power beam width (HPBW) 
for the beams are: $\delta\theta_{3db}^{(1)} = 6.62$ and $\delta\theta_{3db}^{(2)} = 6.56$, respectively.

\begin{figure}[htb]
\begin{minipage}[b]{1\linewidth}
  \centering
  \centerline{\includegraphics[width=7.5cm]{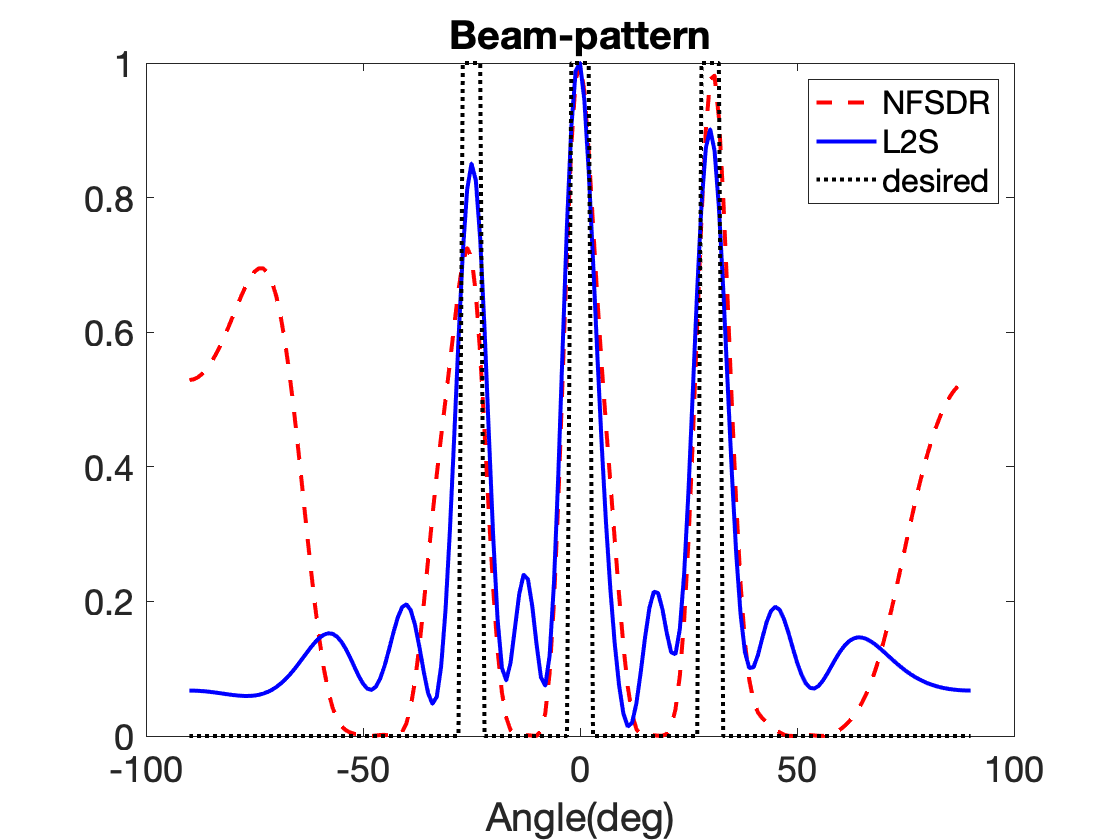}}
  \centerline{\includegraphics[width=9.5cm]{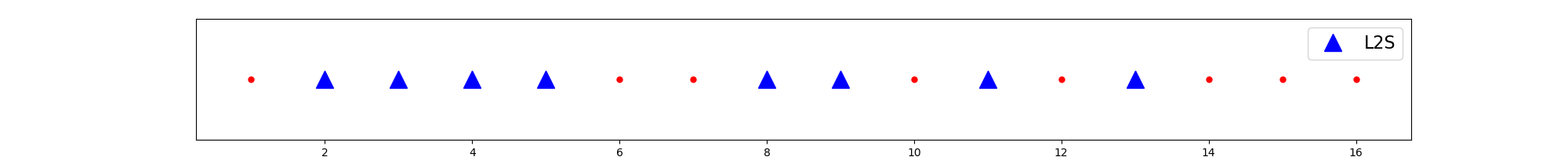}}
  \centerline{\includegraphics[width=9.5cm]{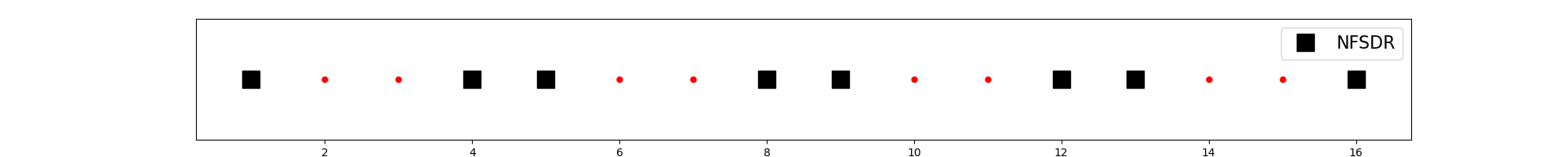}}
\end{minipage}
\caption{Beam-pattern corresponding to the sparse array of $M=8$ elements, selected from a $16$-element ULA  via L2S and NFSDR. Corresponding locations of  selected antennas.}
\label{fig:power3}
\end{figure}

\textcolor{black}{While our proposed antenna selection method was presented for the transmitter side,  it can be easily applied at   the receiver side. We designed an experiment to compare its performance to  the method of \cite{Hamza-Amin-2019}, which also performs antenna selection in order to generate a beam-pattern at certain angles of interest.}
In this experiment the desired beam pattern includes $3$ beams located at $\theta=-25, 0,$ and $30$ degrees. Each beam has width of $5$ degrees. We wish to select $M=8$ antennas from a $N=16$-antenna ULA with  spacing $\lambda/2$. 
The learning parameters are the same as in the previous experiment.
Fig. \ref{fig:power3} shows the obtained power beam-patterns from two methods and the selected antenna locations.
The HPBW for the three beams in our method are: $\delta\theta_{3db}^{(1)} = 8.44$, $\delta\theta_{3db}^{(2)} = 8.8$, and $\delta\theta_{3db}^{(3)} = 9.1$ degrees, respectively. 
\textcolor{black}{It can be seen that the proposed method has better sidelobe control, and this is because it designs the sparse array based on fitting the desired beam-pattern. On the other hand, the method of \cite{Hamza-Amin-2019} focuses only on the lobe at the targeted angle. Moreover, compared to assigning fixed weights on each antenna as in \cite{Hamza-Amin-2019}, our approach to jointly optimize   $\mathbf{Q}$ and $\mathbf{S}$ provides more degrees of freedom, thus resulting in better approximation of  the desired result.}

\begin{figure}[htb]
\begin{minipage}[b]{1\linewidth}
  \centering
  \centerline{\includegraphics[width=7.5cm]{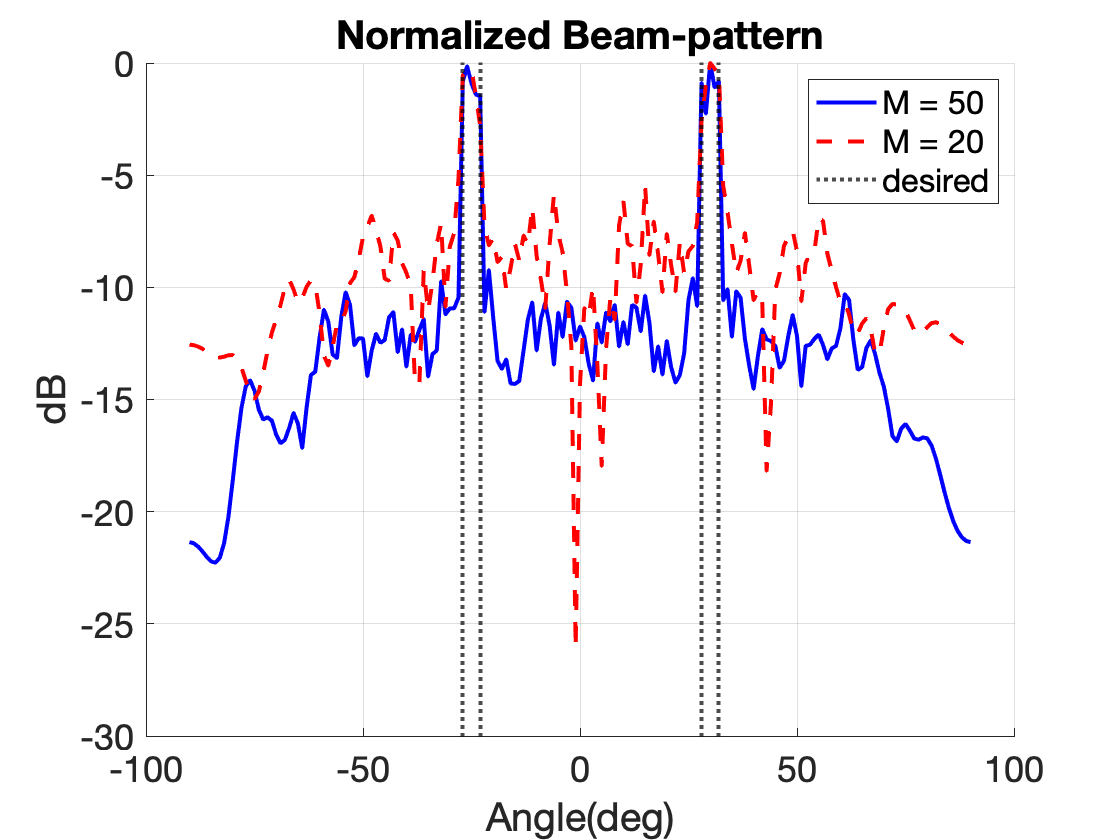}}
\end{minipage}
\caption{Normalized beam-pattern corresponding to the sparse array of $M = 20$ and $50$ elements, selected from a $100$-element ULA  via L2S.}
\label{fig:power4}
\end{figure}

Next, we provide an example of the ability if L2S to work in the case or large arrays. We consider a $100$-element ULA with  half wavelength spacing, and select $M = 20$ or  $M=50$  antennas. The desired power profile is equal to $1$ at angle ranges $[-27,-23]$ degrees and $[28,32]$ degrees, and is zero otherwise. The learning parameters are the same except $T = 300$  and $\alpha$ ranging from $25$ to $125$. The result is shown in Fig.~\ref{fig:power4}.
To get an idea of how high the complexity is when using a  classification method, the number of total classes is ${{20}\choose{100}} > 5\times 10^{20}$. Most 
methods relying on classification \cite{Joung-2016},\cite{Elbir-Mishra-2020},\cite{Vu2019machine} would simply not produce anything for such an array. 
The proposed L2S method took  less than $30$ mins on a Macbook Pro with i9 processors.
{The method of \cite{Hamza-Amin-2019} was also tested but did not produced a result after $4$ hours of running on the aforementioned computer.}

\section{Discussion and Conclusions}

We have proposed L2S,  a novel machine learning approach for antennas selection  and demonstrarted that it can be  used for  efficient design of sparse linear arrays. The method combines multiple coupled Softmax learning models.
L2S has many potential  applications if combined with a suitable loss function. For example, it could be used to solve Knapsack problems, which are combinatorial optimization  problems. 
%
The key advantage of L2S is its complexity, which is $O(N_{epochs}N_{steps}N_{optim})$ proportional to the number of epochs times the number of steps and the complexity of one step of the optimization algorithm. It is not exponential in $M$.




\bibliographystyle{IEEEbib}
\bibliography{sparseantenna,ref_xu,antenna_selection_ML}

\begin{thebibliography}{10}

\bibitem{[12]}
P.~{Stoica}, J.~{Li}, and Y.~{Xie},
\newblock ``On probing signal design for mimo radar,''
\newblock {\em IEEE Transactions on Signal Processing}, vol. 55, no. 8, pp.
  4151--4161, 2007.

\bibitem{[14]}
D.~R. {Fuhrmann} and G.~{San Antonio},
\newblock ``Transmit beamforming for mimo radar systems using signal
  cross-correlation,''
\newblock {\em IEEE Transactions on Aerospace and Electronic Systems}, vol. 44,
  no. 1, pp. 171--186, 2008.

\bibitem{[18]}
H.~{Unz},
\newblock ``Linear arrays with arbitrarily distributed elements,''
\newblock {\em IRE Transactions on Antennas and Propagation}, vol. 8, no. 2,
  pp. 222--223, 1960.

\bibitem{[26]}
N.~{Mitrou},
\newblock ``Results on nonrecursive digital filters with nonequidistant taps,''
\newblock {\em IEEE Transactions on Acoustics, Speech, and Signal Processing},
  vol. 33, no. 6, pp. 1621--1624, 1985.

\bibitem{[28]}
R.~L. {Haupt},
\newblock ``Thinned arrays using genetic algorithms,''
\newblock {\em IEEE Transactions on Antennas and Propagation}, vol. 42, no. 7,
  pp. 993--999, 1994.

\bibitem{[29]}
V.~{Murino}, A.~{Trucco}, and C.~S. {Regazzoni},
\newblock ``Synthesis of unequally spaced arrays by simulated annealing,''
\newblock {\em IEEE Transactions on Signal Processing}, vol. 44, no. 1, pp.
  119--122, 1996.

\bibitem{[30]}
S.~{Holm}, B.~{Elgetun}, and G.~{Dahl},
\newblock ``Properties of the beampattern of weight- and layout-optimized
  sparse arrays,''
\newblock {\em IEEE Transactions on Ultrasonics, Ferroelectrics, and Frequency
  Control}, vol. 44, no. 5, pp. 983--991, 1997.

\bibitem{[32]}
R.~M. {Leahy} and B.~D. {Jeffs},
\newblock ``On the design of maximally sparse beamforming arrays,''
\newblock {\em IEEE Transactions on Antennas and Propagation}, vol. 39, no. 8,
  pp. 1178--1187, 1991.

\bibitem{[33]}
R.~W. Gerchberg,
\newblock ``A practical algorithm for the determination of the phase from image
  and diffraction plane pictures,''
\newblock {\em Optik}, pp. 237--246, 1972.

\bibitem{[35]}
P.~{Stoica}, H.~{He}, and J.~{Li},
\newblock ``New algorithms for designing unimodular sequences with good
  correlation properties,''
\newblock {\em IEEE Transactions on Signal Processing}, vol. 57, no. 4, pp.
  1415--1425, 2009.

\bibitem{[36]}
M.~I. {Skolnik},
\newblock {\em {Introduction to Radar Systems /2nd Edition/}},
\newblock McGraw Hill Book Co., New York, 2 edition, 1980.

\bibitem{[38]}
W.~{Roberts}, H.~{He}, J.~{Li}, and P.~{Stoica},
\newblock ``Probing waveform synthesis and receiver filter design,''
\newblock {\em IEEE Signal Processing Magazine}, vol. 27, no. 4, pp. 99--112,
  2010.

\bibitem{[27]}
P.~{Jarske}, T.~{Saramaki}, S.~K. {Mitra}, and Y.~{Neuvo},
\newblock ``On properties and design of nonuniformly spaced linear arrays
  (antennas),''
\newblock {\em IEEE Transactions on Acoustics, Speech, and Signal Processing},
  vol. 36, no. 3, pp. 372--380, 1988.

\bibitem{Roberts2011}
W.~Roberts, L.~Xu, J.~Li, and P.~Stoica,
\newblock ``Sparse antenna array design for mimo active sensing applications,''
\newblock {\em IEEE Transactions on Antennas and Propagation}, vol. 59, no. 3,
  pp. 846--858, 2011.

\bibitem{Jin_2007}
N.~{Jin} and Y.~{Rahmat-Samii},
\newblock ``Advances in particle swarm optimization for antenna designs:
  Real-number, binary, single-objective and multiobjective implementations,''
\newblock {\em IEEE Transactions on Antennas and Propagation}, vol. 55, no. 3,
  pp. 556--567, 2007.

\bibitem{Schmid_non_uniform_2009}
C.~M. {Schmid}, R.~{Feger}, C.~{Wagner}, and A.~{Stelzer},
\newblock ``Design of a linear non-uniform antenna array for a 77-ghz mimo fmcw
  radar,''
\newblock in {\em 2009 IEEE MTT-S International Microwave Workshop on Wireless
  Sensing, Local Positioning, and RFID}, 2009, pp. 1--4.

\bibitem{Simoni_optimal_array_ICMIM_2018}
M.~A. {González-Huici}, D.~{Mateos-Núñez}, C.~{Greiff}, and R.~{Simoni},
\newblock ``Constrained optimal design of automotive radar arrays using the
  weiss-weinstein bound,''
\newblock in {\em 2018 IEEE MTT-S International Conference on Microwaves for
  Intelligent Mobility (ICMIM)}, 2018, pp. 1--4.

\bibitem{Hamza-Amin-2019}
S.~A. {Hamza} and M.~G. {Amin},
\newblock ``Hybrid sparse array beamforming design for general rank signal
  models,''
\newblock {\em IEEE Transactions on Signal Processing}, vol. 67, no. 24, pp.
  6215--6226, 2019.

\bibitem{Godrich2012}
H.~{Godrich}, A.~P. {Petropulu}, and H.~V. {Poor},
\newblock ``Sensor selection in distributed multiple-radar architectures for
  localization: A knapsack problem formulation,''
\newblock {\em IEEE Transactions on Signal Processing}, vol. 60, no. 1, pp.
  247--260, 2012.

\bibitem{Bahdanau2014neural}
D.~Bahdanau, K.~Cho, and Y.~Bengio,
\newblock ``Neural machine translation by jointly learning to align and
  translate,''
\newblock {\em arXiv preprint arXiv:1409.0473}, 2014.

\bibitem{vaswani2017attention}
A.~Vaswani, N.~Shazeer, N.~Parmar, J.~Uszkoreit, L.~Jones, A.~Gomez,
  {\L}.~Kaiser, and Illia I.~Polosukhin,
\newblock ``Attention is all you need,''
\newblock in {\em Advances in Neural Information Processing Systems}, 2017, pp.
  5998--6008.

\bibitem{Joung-2016}
J.~{Joung},
\newblock ``Machine learning-based antenna selection in wireless
  communications,''
\newblock {\em IEEE Communications Letters}, vol. 20, no. 11, pp. 2241--2244,
  2016.

\bibitem{Elbir-Mishra-2020}
A.~M. {Elbir} and K.~V. {Mishra},
\newblock ``Joint antenna selection and hybrid beamformer design using
  unquantized and quantized deep learning networks,''
\newblock {\em IEEE Transactions on Wireless Communications}, vol. 19, no. 3,
  pp. 1677--1688, 2020.

\bibitem{Vu2019machine}
T.~X. {Vu}, L.~{Lei}, S.~{Chatzinotas}, and B.~{Ottersten},
\newblock ``Machine learning based antenna selection and power allocation in
  multi-user miso systems,''
\newblock in {\em 2019 International Symposium on Modeling and Optimization in
  Mobile, Ad Hoc, and Wireless Networks (WiOPT)}, June 2019, pp. 1--6.

\bibitem{bishop2006pattern}
Christopher~M Bishop,
\newblock {\em Pattern recognition and machine learning},
\newblock Springer, 2006.

\bibitem{Kingma2014adam}
Diederik~P Kingma and Jimmy Ba,
\newblock ``Adam: A method for stochastic optimization,''
\newblock {\em arXiv preprint arXiv:1412.6980}, 2014.

\end{thebibliography}

\end{document}